\documentclass[aps,prl,twocolumn,showpacs,amsmath,amssymb,superscriptaddress]{revtex4}

\usepackage{graphicx}
\usepackage{dcolumn}
\usepackage{bm}
\usepackage{xspace}

\begin{document}


\title{Evidence for muon neutrino oscillation in an accelerator-based 
experiment}

\newcommand{\BCN}{\affiliation{Institut de Fisica d'Altes Energies, Universitat Autonoma de Barcelona, E-08193 Bellaterra (Barcelona), Spain}}
\newcommand{\BU}{\affiliation{Department of Physics, Boston University, Boston, Massachusetts 02215, USA}}
\newcommand{\SACLAY}{\affiliation{DAPNIA, CEA Saclay, 91191 Gif-sur-Yvette Cedex, France}}
\newcommand{\CNU}{\affiliation{Department of Physics, Chonnam National University, Kwangju 500-757, Korea}}
\newcommand{\DU}{\affiliation{Department of Physics, Dongshin University, Naju 520-714, Korea}}
\newcommand{\DUKE}{\affiliation{Department of Physics, Duke University, Durham, North Carolina 27708, USA}}
\newcommand{\KEK}{\affiliation{High Energy Accelerator Research Organization(KEK), Tsukuba, Ibaraki 305-0801, Japan}}
\newcommand{\HIR}{\affiliation{Graduate School of Advanced Sciences of Matter, Hiroshima University, Higashi-Hiroshima, Hiroshima 739-8530, Japan}}
\newcommand{\INR}{\affiliation{Institute for Nuclear Research, Moscow 117312, Russia}}
\newcommand{\KOBE}{\affiliation{Kobe University, Kobe, Hyogo 657-8501, Japan}}
\newcommand{\KOR}{\affiliation{Department of Physics, Korea University, Seoul 136-701, Korea}}
\newcommand{\KYO}{\affiliation{Department of Physics, Kyoto University, Kyoto 606-8502, Japan}}
\newcommand{\LSU}{\affiliation{Department of Physics and Astronomy, Louisiana State University, Baton Rouge, Louisiana 70803-4001, USA}}
\newcommand{\MIT}{\affiliation{Department of Physics, Massachusetts Institute of Technology, Cambridge, Massachusetts 02139, USA}}
\newcommand{\MIYAGI}{\affiliation{Department of Physics, Miyagi University of Education, Sendai 980-0845, Japan}}
\newcommand{\NIIGATA}{\affiliation{Department of Physics, Niigata University, Niigata, Niigata 950-2181, Japan}}
\newcommand{\OKAYAMA}{\affiliation{Department of Physics, Okayama University, Okayama, Okayama 700-8530, Japan}}
\newcommand{\OSAKA}{\affiliation{Department of Physics, Osaka University, Toyonaka, Osaka 560-0043, Japan}}
\newcommand{\ROME}{\affiliation{University of Rome La Sapienza and INFN, I-000185 Rome, Italy}}
\newcommand{\SNU}{\affiliation{Department of Physics, Seoul National University, Seoul 151-747, Korea}}
\newcommand{\SOLTAN}{\affiliation{A.~Soltan Institute for Nuclear Studies, 00-681 Warsaw, Poland}}
\newcommand{\TOHOKU}{\affiliation{Research Center for Neutrino Science, Tohoku University, Sendai, Miyagi 980-8578, Japan}}
\newcommand{\SB}{\affiliation{Department of Physics and Astronomy, State University of New York, Stony Brook, New York 11794-3800, USA}}
\newcommand{\TUS}{\affiliation{Department of Physics, Tokyo University of Science, Noda, Chiba 278-0022, Japan}}
\newcommand{\TRIUMF}{\affiliation{TRIUMF, Vancouver, British Columbia V6T 2A3, Canada}}
\newcommand{\GENEVA}{\affiliation{DPNC, Section de Physique, University of Geneva, CH1211, Geneva 4, Switzerland}}
\newcommand{\UBC}{\affiliation{Department of Physics \& Astronomy, University of British Columbia, Vancouver, British Columbia V6T 1Z1, Canada}}
\newcommand{\UCI}{\affiliation{Department of Physics and Astronomy, University of California, Irvine, Irvine, California 92697-4575, USA}}
\newcommand{\UH}{\affiliation{Department of Physics and Astromony, University of Hawaii, Honolulu, Hawaii 96822, USA}}
\newcommand{\KAM}{\affiliation{Kamioka Observatory, Institute for Cosmic Ray Research, University of Tokyo, Kamioka, Gifu 506-1205, Japan}}
\newcommand{\RCCN}{\affiliation{Research Center for Cosmic Neutrinos, Institute for Cosmic Ray Research, University of Tokyo, Kashiwa, Chiba 277-8582, Japan}}
\newcommand{\VAL}{\affiliation{Instituto de F\'{i}sica Corpuscular, E-46071 Valencia, Spain}}
\newcommand{\UW}{\affiliation{Department of Physics, University of Washington, Seattle, Washington 98195-1560, USA}}
\newcommand{\WARSAW}{\affiliation{Institute of Experimental Physics, Warsaw University, 00-681 Warsaw, Poland}}

\BCN
\BU
\SACLAY
\CNU
\DU
\DUKE
\KEK
\HIR
\INR
\KOBE
\KOR
\KYO
\LSU
\MIT
\MIYAGI
\NIIGATA
\OKAYAMA
\OSAKA
\ROME
\SNU
\SOLTAN
\TOHOKU
\SB
\TUS
\TRIUMF
\GENEVA
\UBC
\UCI
\UH
\KAM
\RCCN
\VAL
\UW
\WARSAW

\author{E.~Aliu}\BCN                               
\author{S.~Andringa}\BCN 
\author{S.~Aoki}\KOBE 
\author{J.~Argyriades}\SACLAY 
\author{K.~Asakura}\KOBE 
\author{R.~Ashie}\KAM 
\author{H.~Berns}\UW 
\author{H.~Bhang}\SNU 
\author{A.~Blondel}\GENEVA 
\author{S.~Borghi}\GENEVA 
% ...
\collaboration{The K2K Collaboration}\noaffiliation

\date{\today}

\begin{abstract}
We present results for $\nu_\mu$ oscillation in the KEK to
Kamioka (K2K) long-baseline neutrino oscillation experiment.
K2K uses an accelerator-produced $\nu_\mu$ beam with a mean energy
of 1.3 GeV
directed at the Super-Kamiokande detector.
We observed the energy dependent disappearance of $\nu_\mu$,
which we presume have oscillated to $\nu_\tau$.
The probability that we would observe these results if there is
no neutrino oscillation is 0.0050\% (4.0$\sigma$). 

\end{abstract}

\pacs{14.60.Pq,13.15.+g,25.30.Pt,95.55.Vj}

\maketitle

\newcommand{\superk}    {Super-Kamiokande\xspace}       

\newcommand{\nue}       {$\nu_{e}$\xspace}
\newcommand{\numu}      {$\nu_{\mu}$\xspace}
\newcommand{\nutau}     {$\nu_{\tau}$\xspace}
\newcommand{\nusterile} {$\nu_{sterile}$\xspace}
\newcommand{\mutau}     {$\nu_\mu \rightarrow \nu_{\tau}$\xspace}
\newcommand{\musterile} {$\nu_\mu \rightarrow \nu_{sterile}$\xspace}
\newcommand{\dms}       {$\Delta m^2$\xspace}
\newcommand{\sstt}      {$\sin^2 2 \theta$\xspace}
\newcommand{\Rnqe}      {$R_{\rm nqe}$\xspace}

\def\EffSKFC{93\%}
\def\NSKexpNoOsc{151^{+12}_{-10}\rm{(syst)}} 
\def\NskFNsyst{5.1\%}
\def\NskNormsyst{5.1\%} 
\def\NSKobs{107} 
\def\NobsIRmu{57} 
\def\NexpIRmu{XX} %
\def\KiiKIaRatio{X.X\%} 
\def\NexpBestfitPhys{103.8} 
\def\SyetEscaleSKi{2.7\%}
\def\SyetEscaleSKii{2.1\%}
\def\ssttBestPhys{1.0} 
\def\dmsqBestPhys{2.8\times 10^{-3}~{\mathrm{eV}^2}} 
\def\ssttBest{1.5} 
\def\dmsqBest{2.2\times 10^{-3}~{\mathrm{eV}^2}} 
\def\KStest{36\%}
\def\KStestPhys{36\%}
\def\KStestNull{0.08\%}
\def\dlnL{9.1}
\def\dlnLNorm{4.3}
\def\dlnLShape{4.5}
\def\NullOsciProbdlnL{0.0050\%(4.0$\sigma$)}
\def\NullOsciProbdlnLNorm{0.26\%}
\def\NullOsciProbdlnLShape{0.74\%}
\def\dmsqNinetyL{1.9}
\def\dmsqNinetyU{3.6}
\def\AtmBG{10^{-3}}

\def\Ln{{\cal L}_{\textrm{num}}}
\def\Ls{{\cal L}_{\textrm{shape}}}
\def\Le{{\cal L}_{\textrm{syst}}}

Recent atmospheric \cite{Ashie:2004mr,Ambrosio:2003yz,Soudan2:2003}, 
reactor \cite{Araki:2004mb}, and 
solar neutrino \cite{Smy:2003jf,Ahmed:2003kj} experiments show that
the existence of neutrino oscillation 
and non-zero neutrino mass are very likely.
Measurements of atmospheric neutrino 
suggest $\nu_\mu$ to $\nu_\tau$ oscillation
with a mass squared difference ($\Delta m^2$) 
around $2.5 \times 10^{-3} \rm{eV^2}$ 
and a mixing angle parameter ($\sin^2 2\theta$) 
that is almost unity \cite{Ashie:2004mr,Fukuda:1998mi}.

The KEK to Kamioka long-baseline neutrino oscillation experiment (K2K)
\cite{Ahn:2001cq,Ahn:2002up} 
is the first accelerator based project to explore neutrino
oscillation in the same $\Delta m^2$ region as atmospheric neutrinos.
The neutrino beam is 
98\% $\nu_\mu$, whose direction is
monitored every beam spill by measuring 
the profile of muons from the pion decays.
The neutrino beam energy spectrum and profile are measured 
by the near neutrino detectors located 300 m from 
the production target. They consists of two detector 
sets: a 1 kiloton water Cherenkov detector (1KT) and a fine grained 
detector system.
The far detector is Super-Kamiokande (SK), a 50 kiloton water 
Cherenkov detector, located 250 km from KEK.

In this letter, we present evidence for the energy-dependent disappearance
of $\nu_\mu$, which are presumed to have oscillated to $\nu_\tau$.
We observe a distortion of the neutrino energy ($E_\nu$) spectrum
and a deficit in the total number of events.  
The expectation for these are derived from measurements
at the near detectors and transformed using
the energy-dependent ratio of the $\nu_\mu$ flux
at the far and near detectors (F/N ratio).
This ratio accounts for the difference between the small portion of
the beam near the center seen by SK and the large section 
of the beam seen by the near detectors.
This is calculated using the neutrino beam Monte Carlo (MC)
simulation and confirmed by measurements of 
pions from the target \cite{Ahn:2001cq,Ahn:2002up}.

{\em{Data sample.}} --- We have analyzed data taken from June 1999 to February 2004, 
which corresponds to 8.9 $\times$ $10^{19}$ protons on target (POT).
From 1999 to 2001 (called K2K-I and SK-I), the inner detector surface
of SK had 11,146 20-inch photo-multiplier tubes (PMTs) 
covering 40\% of the total area \cite{SuperK:2003nim}.
The fine grained detector system was comprised of
a scintillating fiber and water detector (SciFi) \cite{Suzuki:2000nj},
a lead glass calorimeter,
and a muon range detector (MRD) \cite{Ishii:2001sj}.
Starting from January 2003 (K2K-II and SK-II), 19\% of the SK inner detector
is covered using 5182 PMTs, each enclosed in a
fiber reinforced plastic shell with an 
acrylic cover.
The transparency and reflection of these covers in water are 97\% and
1\% respectively.
The near detector data from this period include $2.3 \times 10^{19}$ POT
without the lead glass (K2K-IIa), 
and then $1.9 \times 10^{19}$ POT (K2K-IIb) with a fully-active 
scintillator detector (SciBar) \cite{Nitta:2004nt} in its place.  
Adding the K2K-II data doubles the
statistics compared to the previous analysis \cite{Ahn:2002up}.

The neutrino beam direction is 
monitored using neutrino events in the MRD.
It is stable, within 1 mrad 
throughout the entire experimental period. 
Also, these events confirm that the
energy spectrum is stable.


The 1KT data alone is used to estimate the expected total number of 
events at SK because the 1KT uses the same water target and the 
uncertainties in the neutrino cross section cancel.
The event selection and the 25 ton fiducial volume are the same 
as in \cite{Ahn:2002up}.
We select the subset of events 
in which all the energy is deposited in the inner detector
(fully contained) and only one, muon-like Cherenkov ring is reconstructed
(1-ring $\mu$-like events)
to estimate the $E_\nu$ spectrum along with data from the other near detectors.
For these events, we reconstruct $E_\nu$ by using the measured
muon momentum ($p_\mu$) and direction ($\theta_\mu$).
For the energy spectrum measurement, the largest contribution
to the systematic uncertainty is +2/-3\% in the overall energy scale.


The SciFi detector is made of layers of
scintillating fibers between aluminum tanks filled with water.
The fiducial mass is 5.6 tons.
The K2K-I analysis includes events which reach the MRD and also events
in which the muon track stops in the lead glass, with momentum as low as
400 MeV/$c$, significantly 
lowering the energy threshold compared to \cite{Ahn:2002up}.  
The muon momentum threshold for K2K-IIa 
is 550 MeV/$c$ because in this case we restrict our analysis to events which 
have hit at least two layers in the MRD to improve the purity of muons.


The SciBar detector consists of 14,848 extruded scintillator strips
read out by wavelength shifting fibers and multi-anode PMTs.
Strips with dimensions of $1.3 \times 2.5 \times 300 ~{\rm cm^3}$ 
are arranged in 64 layers. 
Each layer consists of two planes to measure 
horizontal and vertical position.
The scintillator also acts as 
the neutrino interaction target;
it is a fully active detector and has 
high efficiency for low momentum particles.
Although the target is not water, 
possible differences due to nuclear effects are
included in the systematic uncertainty.

In SciBar, tracks which traverse at least three layers ($\sim$8 cm) 
are reconstructed. 
The reconstruction efficiency for an isolated track longer than 10 cm is 99\%.
In the present analysis,
we select charged current (CC) events by requiring 
at least one of the tracks start from the 9.38 ton fiducial volume
and extend to the MRD.
With this requirement, the $p_\mu$ threshold is 450~MeV/$c$.
The $p_\mu$ scale uncertainty, $p_\mu$ resolution, 
and $\theta_\mu$ resolution are 
2.7\%, 80~MeV/$c$, and 1.6$^\circ$, respectively.
The efficiency for a second, short track
is lower than that for a muon track mainly due to 
the overlap with the primary track.
This efficiency smoothly increases from the threshold (8 cm, corresponding 
to 450~MeV/$c$ proton) and reaches 90\% at 30 cm (670~MeV/$c$ for proton).

\begin{table}
  \caption{The reconstruction efficiency [\%] and purity (in parentheses,  [\%]) 
for the quasi-elastic interaction in each sub sample estimated by MC simulation.
}
  \label{table:qeeff}
  \begin{center}
    \begin{tabular}{ccccc} \hline
             & 1-track or& 
             \multicolumn{2}{c}{2-track}  & 
             \hspace*{0.2cm}Total\hspace*{0.2cm} \\
             & 1-ring $\mu$ like & QE & non-QE & \\ \hline
      1KT    & 53 (59) & --- & --- & 53 \\
      SciFi I  & 39 (50) & ~5 (53) & ~2 (11)  & 46 \\
      SciFi IIa & 36 (57) & ~5 (58) & ~2 (12) & 42 \\
      SciBar & 51 (57) & 15 (72) & ~4 (17) & 70 \\
      SK     & 86 (58) & --- & --- & 86 \\ \hline
    \end{tabular}
  \end{center}
\end{table}

For SciFi and SciBar,
we select events in which one or two tracks are reconstructed.
For two-track events, 
we use kinematic information to discriminate between 
quasi-elastic (QE) and non-QE interactions.
The direction of the recoil proton can be predicted
from $p_\mu$ and $\theta_\mu$ assuming a QE interaction.
If the difference between the observed and the predicted direction of the 
second track is within 25$^\circ$, the event is in the QE enriched sample.
Events for which this difference is 
more than 30$^\circ$ (25$^\circ$) for SciFi (SciBar) 
are put into the non-QE sample.
The QE efficiency and purity of the samples are estimated from the 
MC simulation and are summarized in Tab.~\ref{table:qeeff}.


{\em{Near detector $E_\nu$ spectrum.}} ---
We measure the $E_\nu$ spectrum at the near detectors by fitting the 
two-dimensional distributions of $p_{\mu}$ versus $\theta_{\mu}$
with a baseline MC expectation~\cite{Ahn:2002up}.
We simultaneously obtain the cross section ratio
of non-QE to QE interactions ($R_{\rm nqe}$) relative to our MC simulation.
However, we observe a significant deficit of forward going muons in all
near detector data compared to the MC.
To avoid a bias due to this, we perform the $E_\nu$ 
fit using only data with 
$\theta_\mu > 20(10)$ degrees for 1KT (SciFi and SciBar).
The $\chi^2$ value at the best fit is 538.5 for 479 degrees of freedom (DOF).
The resulting $E_\nu$ spectrum and its error are summarized in 
Tab.~\ref{table_enu}, while
the best fit value of $R_{\rm nqe}$ is 0.95.

\begin{figure}
  \begin{center}
    \begin{tabular}{cc}
      \includegraphics[width=0.47\textwidth]{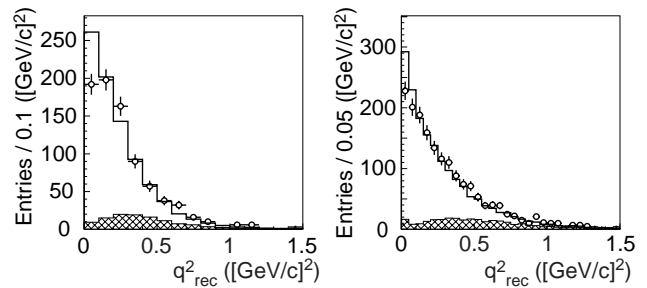}
    \end{tabular}
  \end{center}
  \vspace{-12pt}
  \caption{
  The $q^2_{\rm rec}$ distributions for 2-track non-QE
samples of SciFi (left) and SciBar (right).
Open circles, 
solid lines, and hatched areas show 
data, MC predictions, and CC-QE
component estimated from MC simulation, respectively.
           }
  \label{fig:q2}
\end{figure}

Muons in the forward direction also correspond to events with a 
low value for the square of the momentum transfer ($q^2$),
the relevant parameter in the neutrino interaction models.
From inspection of all subsamples, the amount of resonant pion production
and coherent pion production at low $q^2$ in the MC simulation are possible
sources of the forward muon deficit.
In our MC,
we use the model for resonant pion
by Rein and Sehgal~\cite{Rein:1980wg} with axial vector mass of 1.1 GeV/c$^2$.
For coherent pion, we use the model by
Rein and Sehgal~\cite{Rein:1982pf} with the cross section
calculated by Marteau et al.~\cite{Marteau:1999jp}.
Figure~\ref{fig:q2} shows the $q^2$ distributions 
calculated from $p_\mu$ and $\theta_\mu$ assuming CC-QE kinematics ($q^2_{\rm rec}$).
We modify the MC simulation used in the near and the far detector analysis to 
account for the effect of the observed deficit.
For resonant pions, 
we suppress the cross section by
$q^2/A$ for $q^2<A$ and leave it unchanged for $q^2>A$.
From a fit to the SciBar 2-track non-QE sample, $A$ is
$0.10 \pm 0.03$ (GeV/$c$)$^2$.  
Alternatively,
if we assume that the source of the low $q^2$ deficit is coherent pion production, 
we find the observed distribution is reproduced best 
with zero coherent pion.

\begin{figure}
  \begin{center}
    \begin{tabular}{cc}
      \includegraphics[width=8.0cm]{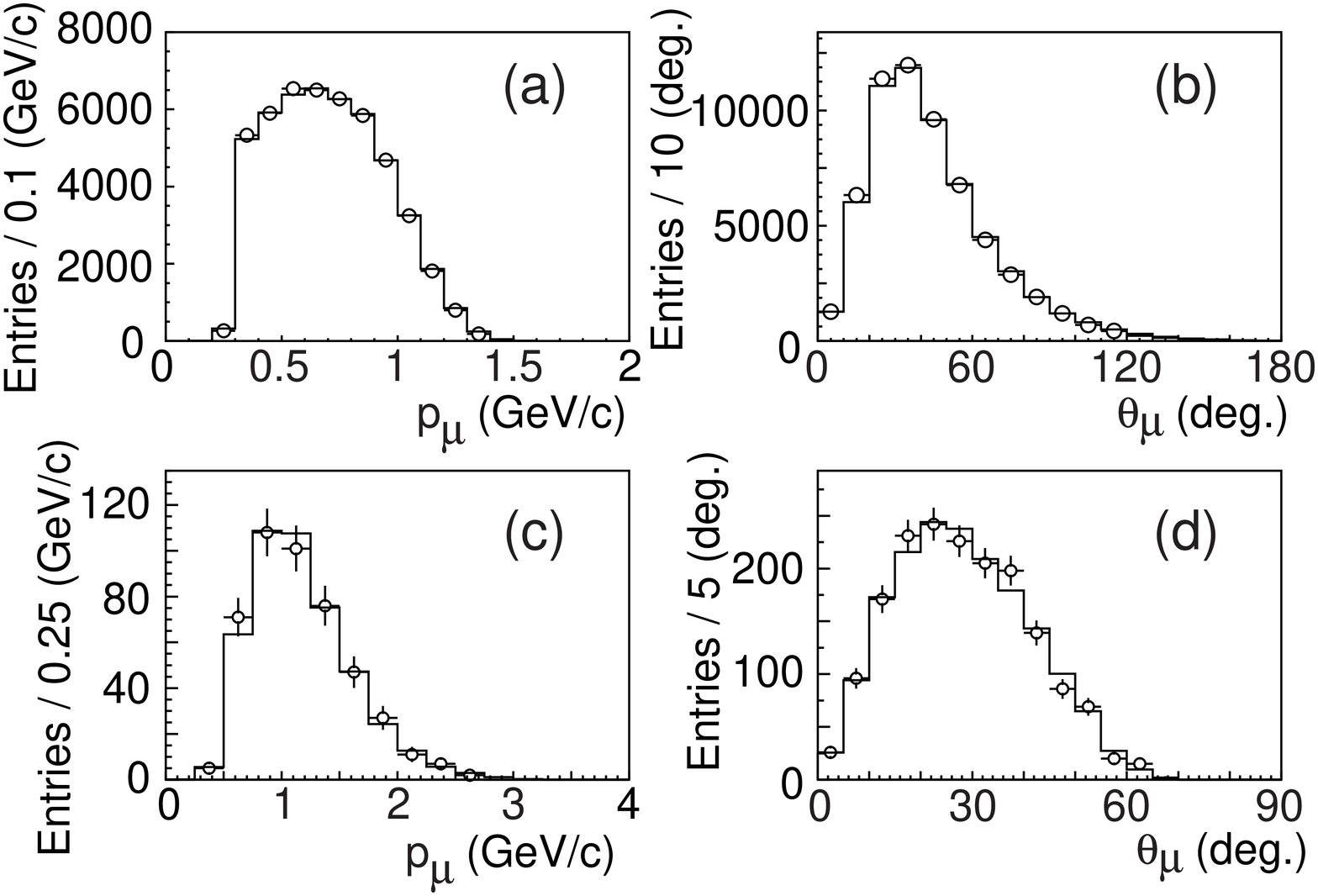} \\
    \end{tabular}
  \end{center}
  \vspace{-12pt}
  \caption{
    A selection of
    muon momentum ($p_\mu$) and direction ($\theta_\mu$) 
    distributions:
    (a) the $p_\mu$ distribution of 1KT fully contained 1-ring $\mu$-like sample,
    (b) 1KT $\theta_\mu$ for the same sample,
    (c) SciFi $p_\mu$ for 2-track QE sample, and
    (d) SciBar $\theta_\mu$ for 2-track nonQE sample.
    Open circles represent data, while histograms are
    MC predictions using the best fit $E_\nu$
    spectrum and suppression of the resonant pion production.
}
  \label{fig:p_mu}
\end{figure}

Considering both possibilities mentioned above,
we fit the parameter \Rnqe again and check the agreement with the data.
The $E_\nu$ spectrum is kept fixed at the values already obtained in the 
first step, but now we use data at all angles.
The best fit value for \Rnqe is 1.02 (1.06)
with $\chi^2$/DOF of 638.1/609 (667.1/606)
when we suppress resonant pion 
(eliminate the coherent pion).
The $p_\mu$ and $\theta_\mu$ distributions from all detectors
are well reproduced for both cases with reasonable $\chi^2$, as
shown in Fig. \ref{fig:p_mu}.
If we repeat the fit with the $E_\nu$ spectrum free, 
the results are still consistent with the first step.
Examining these results carefully, we conclude that we cannot identify
which is the source of the observed deficit in the low $q^2$ region.
Because the value of \Rnqe changes depending on the choice of model,
an additional systematic error of 0.1 is assigned to \Rnqe.
For the oscillation analysis presented in this letter, 
we choose to suppress the resonance production mode in the
MC simulation and when we determine the central value of \Rnqe.
However, we find that the final oscillation results and allowed regions
do not change if we instead choose to eliminate coherent pion,
or use our MC without any corrections. 

\begin{table}
  \caption{The $E_\nu$ spectrum fit results.
    $\Phi_{\rm ND}$ is the best fit value of relative flux
    for each $E_\nu$ bin to the 1.0--1.5 GeV bin.
    The percentages of uncertainties in $\Phi_{\rm ND}$, F/N ratio,
    and reconstruction efficiencies of SK-I and SK-II are also shown.}
  \label{table_enu}
  \begin{center}
    \begin{tabular}{lrcccc} \hline
      \multicolumn{1}{c}{$E_\nu$ (GeV)} & $\Phi_{\rm ND}$ &
      $\Delta(\Phi_{\rm ND})$ & $\Delta(\rm F/N)$ &
      $\Delta(\epsilon_{\mbox{\tiny SK-I}})$ &
      $\Delta(\epsilon_{\mbox{\tiny SK-II}})$ \\ \hline
      $0.0\,~-0.5$  & 0.032 & 46   &  2.6 & 3.7 & 4.5 \\
      $0.5\,~-0.75$ & 0.32 &  8.5 &  4.3 & 3.0 & 3.2 \\
      $0.75-1.0$    & 0.73 &  5.8 &  4.3 & 3.0 & 3.2 \\
      $1.0\,~-1.5$  & $\equiv 1$~{} & --- & 4.9 & 3.3 & 8.2 \\
      $1.5\,~-2.0$  & 0.69 &  4.9 & 10   & 4.9 & 7.8 \\
      $2.0\,~-2.5$  & 0.34 &  6.0 & 11   & 4.9 & 7.4 \\
      $2.5\,~-3.0$  & 0.12 & 13   & 12   & 4.9 & 7.4 \\
      $3.0\,~- {}$  & 0.049 & 17   & 12   & 4.9 & 7.4 \\ \hline
    \end{tabular}
  \end{center}
\end{table}

{\em{Oscillation analysis.}} ---
Events in SK from the accelerator are selected based on timing 
information from the global positioning system.
The background coming from atmospheric neutrinos is estimated to 
be $2 \times 10^{-3}$ events.  For K2K-I+II there are 107 events 
in the 22.5 kiloton fiducial volume that are fully contained, 
have no energy seen in the outer detector, 
and have at least 30 MeV deposited in the inner detector.
The expected number of fully contained events at SK without oscillation is 
$\NSKexpNoOsc{}$.
The major contributions to the errors come from the uncertainties in the
far to near ratio
($\NskFNsyst{}$) and the normalization ($\NskNormsyst{}$);
the latter is dominated by the uncertainty in the fiducial volumes 
due to the vertex reconstruction at both 1KT and SK.

We reconstruct the neutrino energy ($\rm E^{rec}_\nu$),
assuming CC-QE kinematics, 
from $p_\mu$ and $\theta_\mu$ 
for the 57 events in the 1-ring $\mu$-like subset
of the SK data.  With these we measure the energy spectrum distortion
caused by neutrino oscillation.
The detector systematics of SK-I and SK-II are
slightly different because of the change in the number of inner
detector PMTs.  
In the oscillation analysis based on the energy spectrum,
the main contribution to the systematic error
is the energy scale
uncertainty: 2.0\% for SK-I and 2.1\% for SK-II.
Uncertainties for the ring counting and particle identification
are estimated using the atmospheric neutrino data sample and MC simulation.
The differences between the K2K and atmospheric 
neutrino fluxes are also taken into account.


A two-flavor neutrino oscillation analysis, with \numu disappearance,
is performed using a maximum-likelihood method. The oscillation
parameters, $(\sin^22\theta, \Delta m^2)$,  are estimated by
maximizing the product of the likelihood for the observed number of FC
events ($\Ln$) and that for the shape of the  $\rm E^{rec}_\nu$ spectrum ($\Ls$).  The probability density function (PDF) for
$\Ln$ is the Poisson probability for the expected number of events.
The PDF for $\Ls{}$ is the expected $\rm E^{rec}_\nu$ distribution at
SK, which is estimated from the MC simulation.
The PDFs are defined for K2K-I and K2K-II separately. The systematic
uncertainties due to the following sources are taken into account in
the PDFs: the $E_\nu$ spectrum measured by the near detectors, 
the far to near ratio, the reconstruction efficiency and 
absolute energy scale of SK, the ratio
of neutral current to charged current QE cross section, 
the ratio of CC non-QE to
CC-QE cross section and the overall normalization.  
The systematic uncertainties modify the expected distributions,
and each is assumed to follow a Gaussian distribution~\cite{Fukuda:1998mi}.
A constraint term ($\Le$) is multiplied with the likelihood for each of
these systematics, and $\Ln\times\Ls\times\Le$ is maximized during the fit.
The total number of parameters varied in the fit is thirty-four.

\begin{figure}[hbtp!]
  \centering
  \includegraphics[height=5.0cm]{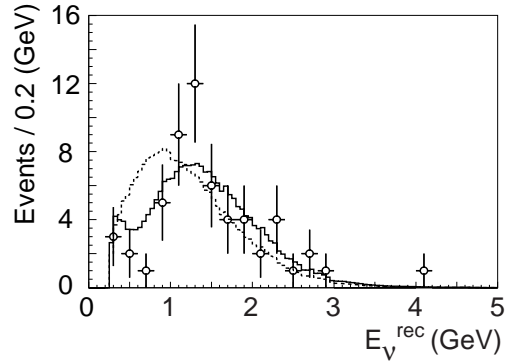}
  \vspace{-10pt}
  \caption{The reconstructed $E_\nu$ distribution
    for the SK 1-ring $\mu$-like sample. 
    Points with error bars are data. 
        The solid line is the best fit spectrum. 
        The dashed line is the expected spectrum without oscillation.
    These histograms are normalized by the number of events observed (\NobsIRmu{}).
        }
  \label{fig:shape}
\end{figure}

\def\ProbAwayFromBoundary{13\%}
\def\SystParamFluctuation{0.9$\sigma$}

The best fit point within the physical region  is  (\sstt,
\dms)=$(\ssttBestPhys{}, \dmsqBestPhys{})$. The expected number of
events at this point is \NexpBestfitPhys{}, which  agrees well with
the 107 observed. The best fit $E_\nu$ distribution is shown
with the data in Fig.~\ref{fig:shape}. 
The consistency between the observed and fit $E_\nu$ distributions
is checked using a Kolmogorov-Smirnov (KS) test.  For the best fit
parameters, the KS probability is \KStestPhys{}, while  that for the
no-oscillation hypothesis is \KStestNull{}. 
The highest likelihood is at a point $(\ssttBest{},
\dmsqBest{})$ which is outside of the physical region.  The probability
that we would get \sstt $\ge$ \ssttBest{} if the true parameters are
our best fit physical parameters is \ProbAwayFromBoundary{}, based on
MC virtual experiments.  For the rest of this letter we refer only to
the physical region best fit. 
The fit results for all the systematic parameters
are reasonable. 
The fits for the K2K-I and K2K-II sub-samples are consistent with the
result for the whole sample.

\begin{figure}[hbtp!]
  \centering
   \includegraphics[height=5.0cm]{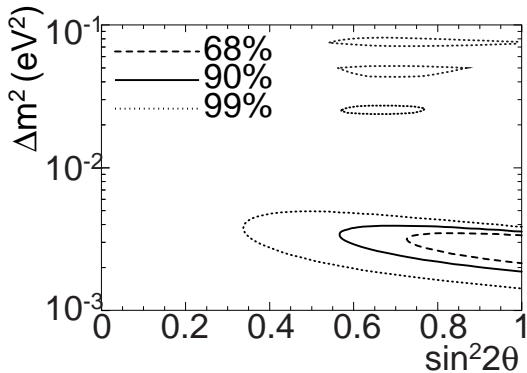}
  \vspace{-10pt}
  \caption{Allowed regions of oscillation parameters. 
     Dashed, solid and dot-dashed lines are 68.4\%, 90\% and
     99\% C.L. contours, respectively.
}
  \label{fig:cont}
\end{figure}

The possibility that the observations are due to a statistical
fluctuation instead of neutrino oscillation is estimated by computing
the likelihood ratio of the no-oscillation case to the best fit point.
If there is no oscillation, the probability of this result is 
\NullOsciProbdlnL{}.
When only normalization (shape) information is used, the probability
is \NullOsciProbdlnLNorm{} (\NullOsciProbdlnLShape{}).  
Allowed regions for the oscillation parameters are evaluated by calculating
the likelihood ratio of each point to the best fit point and are
drawn in Fig.~\ref{fig:cont}.  The 90\% C.L. contour crosses the
$\sin^22\theta=1$ axis at \dms $= \dmsqNinetyL{}$ and  $\dmsqNinetyU{}\times
10^{-3}~\mathrm{eV}^2$.  The oscillation parameters from the 
$E_\nu$ spectrum distortion alone, or the total event analysis alone
also agree. 


In conclusion, 
using accelerator produced neutrinos, we see
the same neutrino oscillation
discovered with atmospheric neutrino measurements.
This result is based on data from 1999 to 2004, 
a total of 8.9 $\times$ $10^{19}$ POT.
The observed number of events and energy spectrum of 
neutrinos at SK are consistent with neutrino oscillation.
The probability that we would see this result if there was
no oscillation is \NullOsciProbdlnL{}.
The allowed regions of the oscillation parameters from the K2K experiment 
are consistent with the atmospheric measurements. 

We thank the KEK and ICRR directorates for their strong support 
and encouragement.  
K2K is made possible by the inventiveness and the
diligent efforts of the KEK-PS machine group and beam channel group.
We gratefully acknowledge the cooperation of the Kamioka Mining and 
Smelting Company.  This work has been supported by the Ministry of 
Education, Culture, Sports, Science and Technology of the Government of Japan, 
the Japan Society for Promotion of Science, the U.S. Department of Energy, 
the Korea Research Foundation, 
the Korea Science and Engineering Foundation,
NSERC Canada and Canada Foundation for Innovation,
the Istituto Nazionale di Fisica Nucleare (Italy),
the Spanish Ministry of Science and Technology, 
and Polish KBN grants: 1P03B08227 and 1P03B03826.

\end{document}